\documentclass{KIArt}
\usepackage[latin1]{inputenc}

\usepackage{amsmath}
\usepackage[english]{babel}
\usepackage{amssymb}

\usepackage[center]{qtree}

\newcommand{\compl}[1]{\overline{#1}}
\def\void{}
\newcommand{\infrule}[3][\void]{%
  {\renewcommand\arraystretch{1.25}
    \ifx\void#1\else#1\hspace{0.5em}\fi
    \begin{array}[c]{@{\hspace*{1em}}c@{\hspace*{1em}}}#2\\\hline #3
    \end{array}}}

\title{Instance Based Methods --- A Brief Overview}
\author{Peter Baumgartner and Evgenij Thorstensen}
\begin{document}
\selectlanguage{english}
\twocolumn[\maketitle
\begin{abstract}%
  Instance-based methods are a specific class of methods for automated
  proof search in first-order logic.  This article provides an overview
  of the major methods in the area and discusses their properties and
  relations to the more established resolution methods.
  It also discusses some recent trends on refinements and applications.

This overview is rather brief and informal, but we provide a
comprehensive literature list to follow-up on the details.
\end{abstract}]

\section{Introduction} 
\emph{Automated Reasoning (AR)} is an application-relevant subfield of
both Logic in Computer Science and Artificial Intelligence. The main
tool is mathematical logic, which provides both a language for
modeling a domain of interest and inference mechanisms for logical
reasoning. AR involves the design of efficient reasoning algorithms
based on logical calculi, their software implementation and
application. AR techniques have been proven useful in areas such as
Computer-Aided Verification, Database Systems, Programming Languages,
Computer Security, Artificial Intelligence, Logic Programming, and
Software Engineering.

In this article we focus on AR in first-order logic (FOL), which has a
long-standing tradition in AI. The perhaps best-known methods are
based on the resolution calculus, which dates back to
1965~\cite{Robinson:65}, or on analytic tableaux methods. (For
example, Prolog's SLD-resolution can be appropriately described as a
tableau method.) With the development of \emph{instance based methods
  (IMs)} since the 1990s, a comparably new family of AR methods for
FOL is available now.  
While also being sound and complete, IMs typically explore a
different search space and exhibit different termination behaviour,
which makes them attractive from a practical point of view as a
complementary method.  For instance, IMs naturally provide decision procedures for
a certain fragment of first-order logic, the Bernays-Sch\"onfinkel
class, which currently attracts a lot of attention as a compact
knowledge representation language.
See also~\cite{Plaisted:TheoremProvingStrategies:CADE:94} for a comparison
of various calculi and strategies, including an instance based
method.  

\section{Ideas and History}
\label{sec:ideas-history}
Like many other AR methods, IMs assume the input formula given as a set of clauses. A
\emph{clause} is an implicitly universally quantified formula of the
form $L_1 \lor \cdots \lor L_n$, where each $L_i$ is a \emph{literal}, i.e.~an
atomic formula or its negation (also known as a \emph{positive} or
\emph{negative} literal, respectively).
The members of a given clause set
are implicitly connected by ``and''. 

As there are efficient translations
from a more general first-order logic syntax into clause
logic~\cite[e.g.]{Nonnengart:Weidenbach:ComputingSmallClauseNormalForms:HandbookAR:2001}, the assumption above does not lead to a loss of generality.
Determining the validity of a given formula---the
most common reasoning problem---translates into a question of 
unsatisfiability of the clause set obtained from the negation of the
given formula (``refutational theorem proving'').

The general idea behind IMs is to prove the unsatisfiability of a
clause set by generating sets of ground instances of its clauses, then
checking the unsatisfiability of these sets. This is a sufficient test
for the unsatisfiability of the given clause set, because, if a set of
instances is not satisfiable then neither is the given clause set.
Consider an example that does this in a very inefficient way. We
generate ground instances (i.e.\ variable-free instances) 
of every clause in the clause set below by looping
over possible constants to try, then check to see, e.g.\ with a
propositional logic SAT solver, whether the generated
instances are unsatisfiable ($x,y,z$ are variables, $a,b$ are constants).
\[
P(x,y)\quad \neg P(a,z) \vee Q(a,z)\quad \neg P(b,z) \vee R(b,z) \quad \neg R(b,c) 
\]
We could start by generating a set of instances
where every variable is replaced by a constant,
 say $a$:
\[
P(a,a)\quad \neg P(a,a) \vee Q(a,a)\quad \neg P(b,a) \vee R(b,a)\quad \neg R(b,c) 
\enspace.
\]
This set is satisfiable, 
so we continue the loop. 
If the enumeration of instances is done in a fair way, we will eventually end up 
with an unsatisfiable set, such as
\[ P(b,c)\quad \neg P(a,c) \vee Q(a,c)\quad \neg P(b,c) \vee R(b,c)\quad \neg R(b,c) \]
and we can stop.

There are two important points of inefficiency here. The first one is
that the clause $\neg P(a,z) \vee Q(a,z)$ is superfluous, as the clause set is
unsatisfiable without it. We should thus avoid generating instances of
this clause altogether. The second point of inefficiency is the lack
of direction in the instantiation process, leading to many dead-ends
as we generate instances that do not lead to an unsatisfiable
set. Usually, the dead-ends can be avoided by keeping the generated sets, but one still has to deal with all the superfluous
instances generated. Contemporary IMs try to avoid these and other
problems in order to obtain practically useful methods.

The idea explained above is already present in the work by Davis,
Putnam, Logemann and Loveland, and others, in the early sixties of the
last
century~\cite{Davis1960a,Davis:Loveland:Logeman:DPLL:62}.
One of their algorithms to check the satisfiability of the sets of instances is still popular as
the basis of modern propositional SAT solvers, commonly
referred to as the ``DPLL procedure''. However, because of its
primitive treatment of quantifiers by uninformed guessing, their
first-order logic procedures have quickly been overshadowed by a
reasoning procedure developed soon afterwards, the \emph{resolution
  calculus}~\cite{Robinson:65}.  

One of the key insights in~\cite{Robinson:65}
concerns the use of MGUs (most general unifiers) for reasoning 
on clauses, as in the \emph{resolution
  inference rule}:\footnote{It took another 25 years until
the development of the ``modern'' theory of resolution had begun in the
1990s~\cite{Bachmair:Ganzinger:90}. This lead to a breakthrough in
resolution theory by unifying more or less all resolution variants and
improvements until 
then in a single theoretical framework, yet more elegant, general and
powerful~\cite{Bachmair:Ganzinger:ResolutionTheoremProving:Handbook:2001}.
}
\begin{displaymath}
\infrule{
    C \lor K \qquad L \lor D
}{
    (C \lor D)\sigma
}\quad\mbox{if $\sigma$ is a MGU of $K$ and $\compl{L}$}
\end{displaymath}
That $\sigma$ is a \emph{most general} unifier means that, given any
unifier $\tau$ (a substitution that makes the involved literals
equal), there is  a substitution $\gamma$ such that $\sigma\gamma = \tau$.
The notation $\compl L$ refers to the complement of $L$.

In contrast to resolution, IMs work with instances of clauses without combining them
 into new clauses as the resolution inference rule does. 
With  that view, bottom-up model generation procedures like 
SATCHMO~\cite{Manthey:Bry:SATCHMO:88} and hyper
tableau~\cite{Baumgartner:Furbach:Niemelae:HyperTableau:JELIA:96}
qualify as IMs. They employ unification at the core of their
inference rules, but still require blind guessing of ground instances
in certain circumstances.
A stream of research that avoids this was initiated with 
Lee and Plaisted's first IM, the Hyper-Linking calculus
\cite{Lee:Plaisted:92}. Hyper-Linking is akin
to the instantiation loop described above, but uses unification to
guide the instantiation of clauses and capitalizes on the latest SAT
solver technology for the unsatisfiability check of the instance sets. Since
then, other IMs have been developed by Plaisted and his coworkers
\cite{Chu:Plaisted:SematicHyperLinking:CADE:94,Plaisted:Zhu:OrderedSemanticHyperLinking:JAR:00}.
Another influential approach is Billon's disconnection
calculus~\cite{Billon:DisconnectionMethod:TABLEAUX:96}, which was
picked up by Letz and Stenz and has been significantly developed
further since then into a tableaux-like
IM~\cite{Letz:Stenz:ModelGenerationDisconnectionTableaux:LPAR:2001,Letz:Stenz:GeneralizedVariablesDisconnection:IJCAR:04}. 
To some extent, the
tableau structure of Disconnection Tableaux enables one to avoid the problem of  
superfluous clauses.

One of the authors of this paper introduced a first-order version of
the propositional DPLL procedure mentioned above,
First-Order DPLL (FDPLL)~\cite{Baumgartner:FDPLL:CADE:00},
which is now subsumed by the Model Evolution (ME)
calculus~\cite{Baumgartner:Tinelli:ModelEvolutionCalculus:CADE:2003}. 
These two calculi are less concerned with generating sets of instances
of the clauses themselves; instead they focus on finding a potential model
for the clause set by a semantic-tree construction. The 
model representation formalism in the ME calculus has been studied in
its own right by~\cite{Fermueller:Pichler:Contexts:CADE:2005}. 

Other IMs have also been described
in~\cite{Baumgartner:HyperNextGeneration:Tableaux:98,Baumgartner:Eisinger:Furbach:CCC:CADE:99},
by Hooker~\cite{Hooker:etal:PartialInstantiationMethods:JAR:2002}, and
by Ganzinger and
Korovin~\cite{Ganzinger:Korovin:NewDirectionsInstanceBased:LICS:2003}.
See~\cite{Jacobs:Waldmann:ComparingIM:JAR:2007} for a thorough
comparison of some of these IMs.  A rather recent development
is~\cite{Moura:Bjoerner:EPRUsingDPLL:IJCAR:2008}, which employs
tuple-at-a-time reasoning for sets of ground instances of clauses,
which are represented by BDD.

In the following we describe the idea behind some of these methods in more detail. A
key notion to several of them is that of a \emph{link} between two
clauses. Given two clauses $C \vee K$ and $L \lor D$, the literals $L$ and $K$ constitute a
\emph{link} if there is an MGU $\sigma$ for $L$ and $\compl K$. 

\section{Inst-Gen}
\label{sec:inst-gen}
Inst-Gen~\cite{Ganzinger:Korovin:NewDirectionsInstanceBased:LICS:2003}
is perhaps the conceptually simplest IM. Unlike the unguided
IM in Section~\ref{sec:ideas-history}, Inst-Gen uses 
unification to generate clause instances, which is realized with the
following inference rule:
\begin{displaymath}
\infrule{
    \hphantom{(}C \lor K\hphantom{)\sigma} \qquad \hphantom{(}L \lor D\hphantom{)\sigma}
}{
    (C \lor K)\sigma \qquad (L \lor D)\sigma
}\quad\mbox{if $\sigma$ is a MGU of $K$ and $\compl{L}$}
\end{displaymath}
Notice that the side condition 
in the rule can be replaced by ``$L$ and $K$ constitute a link with unifier $\sigma$''.
The rule can be strengthened by requiring that at least one conclusion clause
must be a \emph{proper} instance of its premise, that is, the conclusion must
be an instance of its premise but not the other way round.

The Inst-Gen inference rule differs from the resolution rule by
keeping the instantiated premises separate instead of combining them
into one resolvent. While resolution derives the empty clause $\bot$ to
indicate unsatisfiability, Inst-Gen uses a SAT solver to check
propositional unsatisfiability of the clause set after instances have
been added. For that, every variable in every clause is uniformly
replaced by the same constant. Intuitively, 
by this process an unsatisfiable clause set gets closer and closer to being
propositionally unsatisfiable. 

The beauty of Inst-Gen is that it is ''trivially'' sound and not so
difficult to prove complete. It is sound because it always adds
instances of clauses already present, hence consequences thereof.
To get a deeper understanding of how the calculus works it is
instructive to sketch its completeness proof. 
As a prerequisite for that, let $\bot$ denote
the substitution already mentioned above that uniformly replaces every variable
by some fixed constant. Also, assume as given a clause set $M$ that is
closed under the application of the Inst-Gen inference rule, modulo
alphabetic variants of clauses.
Completeness then amounts to showing that 
if $M$ is unsatisfiable then $M\bot$ is likewise unsatisfiable. 
It is advantageous, however, to work in the contrapositive
direction. Thus assume that $M\bot$
is satisfiable; it suffices to construct a model for $M$.

Because $M\bot$ is satisfiable, there is a satisfiable \emph{path} $P$
through $M\bot$, that is, a set of literals that is obtained by
picking exactly one literal from each clause in $M\bot$ and such that
$P$ does not contain a link, i.e., two complementary literals.

We sketch how $P$ guides the construction of a model $I$ for $M$,
better said, a partial model for $M$ that can be extended to a total
one (a partial model is a consistent set of ground
literals): initially let $I$ be the empty set. For every clause $C \in
M$ and every ground instance $C\gamma$ do the following: if
\begin{enumerate}
\item  $C$ is the
\emph{most specific representation} of $C\gamma$, that is, there is no
clause $D \in M$ that is a proper instance of $C$ such that $C\gamma =
D\gamma'$ for some $\gamma'$, and
\item $C\gamma$ is false in $I$, and
\item $K$ is a literal in $C$ such that $K\bot \in P$ and 
  $K\gamma$ is undefined in $I$ (i.e., neither $K\gamma$
nor its complement $\compl{K}\gamma$ are in $I$)
\end{enumerate}
then add $K\gamma$ to $I$. In this case we say that $C\gamma$ \emph{generates}
$K\gamma$ (in $I$).

Now, assume to the contrary that $I$ falsifies some ground instance
$D\gamma$ of a clause $D \in M$. We show that ``enough'' inferences must have
been applied so that $I$ satisfied $D\gamma$, this way contradicting the
assumption.  Without loss of
generality assume that $D$ is a most specific representation of $D\gamma$
(among all representations there is always a most specific one). Let
$L \in D$ be a literal such that $L\bot \in P$ (recall that $P$ contains a
literal from every clause in $M\bot$, thus we can find such an $L$).  As
$D\gamma$ is false in $I$, $D\gamma$ cannot be generating and we must have
$\compl{L}\gamma \in I$. That is, some clause $C\gamma$ generates $\compl{L}\gamma$ in
$I$. Let $K\in C$ be a literal such that $K\gamma = \compl{L}\gamma$.  Now, the
Inst-Gen inference rule is applicable to $C$ and $D$ by using $K$ and
$L$ as the link literals. Let $\sigma$ be the MGU of $K$ and
$\compl{L}$. By condition 3 above it holds that $K\bot \in P$. As $L\bot \in P$ the
MGU $\sigma$ must replace at least one variable in $K$ or in $L$ by a
non-variable term, otherwise $P$ would contain complementary
literals. Therefore, $C\sigma$ is a proper instance of $C$, or $D\sigma$ is a
proper instance of $D$ (or both). Both cases give a contradiction: in
the first case, $C\gamma$ would not be generating as
$C\sigma$ is a more specific representation of $C\gamma$ (recall that $M$ is
closed under Inst-Gen, and hence $C\sigma \in M$). Similarly, in the second
case $D$ would not be a most specific representation of $D\gamma$ by virtue
of $D\sigma \in M$. Thus, the assumption that $I$ falsifies a $D\gamma$ cannot hold,
and the proof is complete.

The Inst-Gen calculus is extensible by many refinements, including
hyper-style inference, semantic selection (add instances that are not
yet satisfied by the propositional model $P\bot$), and certain forms of
redundancy elimination, which can all be justified by the
model-generation completeness proof above.

\section{Hyper-Linking}
The Hyper-Linking
calculus~\cite{Lee:Plaisted:92} works,
similarly to Inst-Gen, by looking 
for links between clauses. The process is divided into ``rounds'';
each round computes for every clause $C$ the set of hyper-links for
this clause. A hyperlink for $C = L_1 \vee \cdots \vee L_n$ is a set of links
between $C$ and other clauses such that for every $i$, there is one
and only one link featuring $L_i$.

For every such hyper-link, one can compose the unifiers of every link
in it (possibly renaming variables) to get a single substitution for
the hyper-link. Call this substitution $\theta$; the instance added to the current
clause set is $C\theta$. After all the clauses have been processed (this
process terminates), the calculus temporarily 
grounds every clause by substituting some
distinct constant for every variable to obtain a set of ground
clauses, as for Inst-Gen. This set can be checked for satisfiability by
e.g.~resolution, and if it is unsatisfiable the procedure stops. Otherwise, a new
round is initiated with the set of instances that the last round failed to refute as
input.

In this calculus, the instances computed are an attempt to create complementary clauses.
 This helps to avoid superfluous clauses, as
they would have few or no links to other clauses. However, if there is
a ``cluster'' of superfluous clauses, i.e.~clauses that have links
between them but no links to other clauses, and this cluster is
satisfiable, Hyper-Linking would still generate instances of clauses
in the cluster.

The Hyper-Linking calculus has been developed further, and has lead to
IMs that take advantage of semantic guidance and of certain
refinements based on ordering
restrictions~\cite{Chu:Plaisted:SematicHyperLinking:CADE:94,Plaisted:Zhu:OrderedSemanticHyperLinking:JAR:00}.

\section{Disconnection Tableaux}

The Disconnection Tableaux
calculus~\cite{Letz:Stenz:ModelGenerationDisconnectionTableaux:LPAR:2001}
works on links, which are also called connections in this context.  In
contrast to Hyper-Linking, where all the links of a clause are used
together, Disconnection looks at one link at a time. The general idea
is as follows: One starts by setting up an \emph{initial path} through
the input clause set by arbitrarily picking one literal from each
input clause. This initial path is fixed throughout the entire
subsequent tableau construction.  Starting with the initial path, the
calculus' single inference rule takes a branch constructed so far and
looks for a link between two literals on the branch. If such a link
exists and, say, $C$ and $D$ are two clauses containing the link
literals, the calculus expands the branch by the clause instances $C\sigma$
and $D\sigma$ using the unifier $\sigma$ for this link. This way, the link is
``disconnected''.  Every relevant instance that can be found this way
eventually ends up on the tableau, and if a branch contains literals
that become complementary when grounded to a special constant, it is
closed. A proof is a tableau where every branch is closed.

Consider an example expansion with the clause $\neg P(a,z) \vee Q(a,z)$, with $\neg
P(a,z)$ being the literal on the initial path to the following tableau
(the initial path is not drawn):

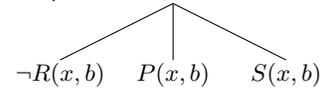
\Tree
        [.{}
          [.{$\neg R(x,b)$}
          ]
          [.{$P(x,b)$}
          ]
          [.{$S(x,b)$}
          ]
        ]
\medskip

To use the link between $P(x,b)$ and $\neg P(a,z)$, the variable $x$
has to be instantiated to $a$, and $z$ to $b$. 
Before doing the expansion with $\neg P(a,z) \vee Q(a,z)$,
the instantiated tableau clause has to be put on the branch. The branch
marked by $\star$ is closed, as it contains complementary literals $P(a,b)$
and $\neg P(a,b)$.

After the expansion, the tableau looks like this:
  
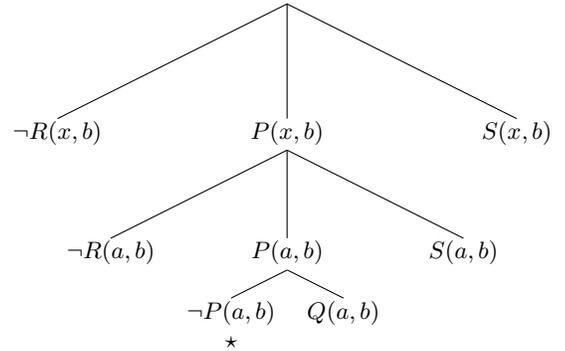
\Tree
        [.{}
          [.{$\neg R(x,b)$}
          ]
          [.{$P(x,b)$}
            [.{$\neg R(a,b)$}
                ]
                [.{$P(a,b)$}
                  [.{$\neg P(a,b)$\\
                         $\star$}
                  ]
                  [.{$Q(a,b)$}
                  ]
                ]
                [.{$S(a,b)$}
                ]
          ]!\qsetw{4.5cm}
          [.{$S(x,b)$}
          ]
        ]
\medskip

In general, the test for complementary literals is carried out 
after the tableau has been temporarily grounded by
replacing every variable by a constant, as for Inst-Gen and Hyper-Linking.
Any branch satisfying this
condition without looking at the initial path can be closed, and there is always at least one such branch in every
expansion. 

This way of looking for single links that are ``needed'', i.e.~can be
used together with the clauses that are already there, removes the problem of
superfluous clauses to a large extent, as they have few or no links to
other clauses. 
Hyper-Linking, for example, does not have a way to avoid
this possibility. On the negative side, Disconnection generates many
similar branches, something that can lead to near-copies of
derivations.

\section{FDPLL and Model Evolution}
\label{sec:fdpll-me}
The First-Order DPLL (FDPLL)~\cite{Baumgartner:FDPLL:CADE:00} calculus
and its successor, Model Evolution
(ME)~\cite{Baumgartner:Tinelli:ModelEvolutionCalculus:CADE:2003} share
some features with the calculi described above. The main object of a
derivation is a tree, as in Disconnection Tableaux, but branching is
on complementary literals instead of clauses. As in Hyper-Linking,
every literal of the current clause must be simultaneously linked with literals on
the current branch to drive the inference rules. 

FDPLL and ME have been introduced as a lifting of the propositional
core of the DPLL procedure to the first-order level. To describe how they work, it is
instructive to recapitulate propositional DPLL first: Given a
propositional clause set $S$, one picks a propositional variable, say
$A$, from a clause in $S$, and creates two new clause sets $S[A/\top]$
and $S[A/\bot]$ (by $S[A/\bot]$ we mean the set $S$ with every instance of
$A$ replaced by $\bot$), to be analyzed separately. The two clause sets
created are simpler than $S$, and can be further simplified by
propositional rules, e.g.~$A \vee \top \equiv \top$. If we find an elementary
contradiction during this simplification, that clause set is
unsatisfiable. If not, a new propositional variable to split on is
picked, until all propositional variables have been exhausted (with
the conclusion that $S$ is satisfiable), or all the sets generated are
shown to be unsatisfiable, which means that $S$ is not satisfiable
either.

FDPLL lifts this splitting rule to first-order clauses by 
case analysis on possibly non-ground literals. FDPLL can be described
as a semantic tree construction method akin to
Disconnection tableaux, but branching on complementary literals
instead of clauses. The intention of the calculus is then to construct
a model for the input clause set. As an example, consider the two clauses
\[P(a,y) \qquad P(x,b) \vee \neg P(z,y) \vee Q(x,y,z)
\]
and suppose that the following tree has already been derived:
\medskip
  
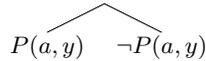
\Tree
      [. 
        [.{$P(a,y)$}
        ]
        [.{$\neg P(a,y)$}
        ]
      ]
\medskip

The left branch specifies the interpretation that assigns true to
all ground instances of $P(a,y)$, and false to all other atoms.
This interpretation assigns false to the second clause --- in fact,
it assigns false to every instance of $P(x,b) \vee \neg P(a,y) \vee
Q(x,y,a)$ where $x$ is different from $a$. This situation is detected by unifying the branch
literal $P(a,y)$ with the complement of the clause literal $\neg P(x,y)$. Such
links with branch literals have to be found for every clause
literal. Positive clause literals can also be linked with 
the pseudo-literal ``$\neg x$'' in the root node, where $x$ unifies with any
positive literal. 
In general, if a clause instance $C$ computed this
way contains a literal $L$ such
that neither $L$ nor the complement of $L$ is on the branch, the
branch is extended by splitting on $L$ and its complement. 
In the example, the literal $P(x,b)$ can be used for this, but $\neg
P(a,y)$ cannot. On the left branch
the interpretation will at least partially be ``repaired'' towards a
model for $C$, and on the right branch the clause $C$ is implicitly
shrunk by the presence of the complement of $L$. Both cases mark some
progress in either finding a model or a refutation. The test for closing
branches is similar to the one in Disconnection Tableaux, and is based on
simultaneously unifying all the literals of a clause with complementary branch
literals after temporarily grounding them. 

FDPLL is sound and complete. Regarding soundness recall that branch
closure is based on temporarily replacing every variable in every
branch literal by some fixed constant. A closed tree can therefore
be seen as one that branches on complementary propositional
literals. Soundness of FDPLL then follows easily.

The completeness proof is of the model-generating kind, akin to the
one for Inst-Gen (cf.\ Section~\ref{sec:inst-gen}). In FDPLL, the
central concept is that of a \emph{candidate model} induced by a
(current) branch. Similarly to Inst-Gen, one can argue that any
open branch in a fair derivation has had ``enough'' inferences applied 
to it, so no falsified clause can exist.  We just note
one key property here: Whenever the candidate model falsifies a clause
$C$, there is a clause instance $C\sigma$, where $\sigma$ simultaneously
unifies the links between all literals in $C$ and branch literals. The branch is then either closed or the branch can be split with some
literal in $C\sigma$. Conversely, if no such unifier $\sigma$ exists then $C$
holds true in the candidate model. This fact can be seen as a
semantically justified redundancy criterion -- inferences with ``true''
clauses need not be carried out.

The following semantic tree can be derived from the example tree above in two steps:
\medskip
  
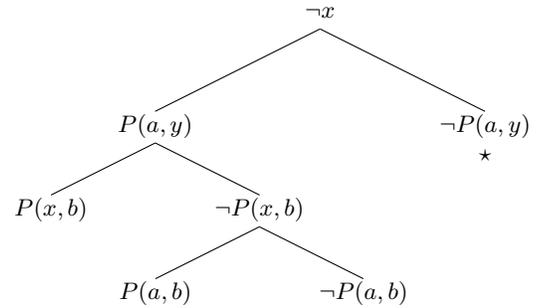
\Tree
      [.{$\neg x$}
        [.{$P(a,y)$}
                        [.{$P(x,b)$}
                        ]
                        [.{$\neg P(x,b)$}
                                [.{$P(a,b)$}
                                ]
                                [.{$\neg P(a,b)$}
                                ]!{\qbalance}
                        ]
        ]
        [.{$\neg P(a,y)$\\
                   $\star$}
        ]
      ]
\medskip

The ME calculus subsumes FDPLL by also lifting several simplification
rules from DPLL to the first-order level. ME operates
on sequents of the form $\Lambda \vdash \Phi$. The set of literals $\Lambda$ corresponds
to a branch in an FDPLL semantic tree and is called a context, while $\Phi$ is
the current clause set to be refuted. 
As said earlier, the chief advantage of ME over FDPLL is that it
manages to lift the simplification rules of DPLL to first-order
logic. An example of this is the Subsume rule, which in the
propositional case allows one to simplify a clause set $\{L, L \vee C, \ldots\}$ to $\{L, \ldots\}$
--- as $L$ has to be true, the satisfiability of this clause set does
not depend on $L \vee C$. A dual rule, Unit Resolution, allows to remove a literal from a
clause in the presence of a unit clause with a complementary literal.
Such rules are mandatory in practice, and ME contains first-order
logic variants of both of them. They are used to simplify the current
clause set $\Phi$ based on the current context $\Lambda$.

In addition to the simplification rules there are two rules that add
literals to the context, Assert and Split. The Split rule is similar
to the splitting rule in FDPLL, while the Assert rule is a lifting of
the propositional rule that assigns $\top$ to a clause containing a
single literal, called a \emph{unit clause}. Such a clause can only be
made true in one way, namely by assigning true to the literal (in the
propositional case), or assigning true to every instance of the
literal (in the first-order case).

\section{EPR}
In the previous sections we have reviewed several IMs by summarizing
their main underlying ideas and differences. IMs are conceptually
different to more established methods based on resolution or 
unification-based free-variable tableaux. This way, they contribute as
alternative methods to the repository of available first-order AR
methods. In particular, IMs are strong on a
certain fragment of first-order logic that proves difficult for many other methods. More
precisely, all instance-based methods can be used as decision
procedures for 
the Bernays-Sch\"onfinkel (BS) class of first-order logic. This class
consists of formulas that, when written in prenex normal form, have
the form $\exists\vec{x}\forall\vec{y}\phi(\vec x, \vec y)$, where $\phi$ is a
quantifier-free formula without function symbols. Such formulas are sometimes also 
referred to as effectively propositional (EPR), since they can be
effectively translated into propositional logic by a finite
process of grounding. However, the cost of that is an exponential blow-up in
formula size. The translation into clause logic always yields
a clause set that may contain variables and constants, but no terms
built from proper function symbols (``Datalog''). 

This property helps to explain why IMs decide the satisfiability
problem of the EPR class. It is simply
because the set of instances of a finite clause set without function symbols is finite 
modulo renaming of clause variables, something that is easy to control
in IMs.
By contrast, resolution (see the resolution inference  rule in
Section~\ref{sec:ideas-history}) might derive clauses
of unbounded length, which makes it less suitable for EPR.
This circumstance may partially explain why the winners in the EPR
category in the annual CADE ATP Systems competition
(CASC~\cite{SS06-SoCASC}), a 
major competition for automated theorem provers, have for the
last six years been instance-based provers instead of
resolution-based provers. 

The decidability problem of EPR is complete for NEXPTIME. In practical
terms this means that much more succinct problem specifications are
possible with EPR than with propositional logic.  This suggests to
capitalize on IMs as decision procedures for EPR and to investigate
practically feasible reductions of application problems into EPR.  For
instance, it is already known that bounded model checking problems can
be encoded in BS logic much more succinctly than in propositional
logic~\cite{Perez:Voronkov:BoundedModelCheckingEPR:CADE:2007}.

Another example is the optimized functional translation of modal
logics~\cite{OhlbachSchmidt97} to BS
logic~\cite{Schmidt:DecidabilityResolutionModal:JAR:99}. Many
benchmark 
problems obtained this way are contained in the TPTP problem
library~\cite{Sutcliffe:Suttner:TPTP:JAR:98}, and implementations of
instance based methods consistently score very well on them.  In the
description logic context, \cite{DBLP:conf/semweb/MotikSS04} show how
to translate the expressive description logic ${\cal
  SHIQ}(\mathrm{D})$ to BS logic, but with a different motivation.

Other potentially useful applications of IMs as
decision procedures for EPR
lie within the \emph{constraint programming} area.  IMs are possibly not 
the preferred choice as solvers for search problems, typically in NP, as
this is the domain of the traditional constraint programming paradigm. More
appropriate seems the application to e.g. model expansion
problems~\cite{U-SFraser-CMPT-TR:2006-24} (with NEXPTIME combined
query/data complexity), which can be reduced to EPR in a way similar
to finite model computation mentioned above. 
Another application is to analyze constraint
models for certain ``interesting'' properties, like proving of 
functional dependencies and
symmetries between decision variables~\cite{Cadoli:Mancini:TheoremProver:AIIAL2005,Cadoli:Mancini:ExploitingFunctionalDependencies:JELIA:2004}. Quite
often, the resulting proof obligations lie within BS logic.  

A ``generic'' application area is \emph{finite model
  computation}. 
Finite model computation is the problem of computing a
model with a finite domain for the given formula or clause set, if one
exists. One application of this is in computing counterexamples
of ``false'' theorems, which arise frequently in software verification
or modelling in early
stages. See~\cite{Bry:Torge:DeductionMethodFiniteSatisfiability:JELIA:98,Baumgartner:Schmidt:BUMGEnhanced:IJCAR:2006,deNivelle:Meng:GeometricResolution:IJCAR:2006}
for IM-related methods. Other methods for finite model computation
essentially work by stepwise reduction to formulas in propositional
logic and use a SAT solver on the result.~\cite[e.g.]{Slaney:FINDER:CADE:94,McCune:MACE:94,Zhang:SEM:IJCAI:95,Claessen:Soerensson:MACEimprove:ModelComputationWS:2003,conf/tableaux/Peltier03}.
In~\cite{Baumgartner:etal:MEFiniteModels:JAL:2009} it was
shown how this model computation paradigm can be rooted in
the Model Evolution calculus instead of a SAT solver, something that can lead
to space advantages. Actually, the results
in~\cite{Baumgartner:etal:MEFiniteModels:JAL:2009} are more general,  
and any method that decides the EPR class can be used.

\section{Conclusions and Outlook}
In this paper we surveyed methods for instantiation-based theorem
proving and indicated their strength for the EPR class of first-order
logic formulas. 
We concentrated on the basic versions of four typical IMs, Inst-Gen,
Hyper-Linking, Disconnection Tableaux and FDPLL/Model Evolution.
The theoretical power of each of them is the same, they all are sound
and complete methods for first-order clausal theorem
proving. So what are the differences? 
First, there are conceptual differences in the way they lift
propositional reasoning to the first-order level. IMs can broadly be classified
as \emph{one-level} vs.\ \emph{two-level methods}: two-level methods
like Inst-Gen and Hyper-Linking directly employ a propositional
SAT-solver as a subroutine for 
periodic unsatisfiability tests of the grounded version of the current
clause set. This way, these methods can always capitalize on the latest advances in
SAT-solving technology. In the extreme case, when the given clause set
is already ground, their performance is the same as if their
SAT-solver had been directly called. In contrast, one-level
methods like Disconnection Tableaux and FDPLL/Model Evolution work directly 
with ``lifted'' first-order logic data structures and inference rules of their 
propositional base calculi --- in these cases propositional tableaux and propositional
DPLL. These first-order data structures allow some
optimizations that are difficult to replicate on the
propositional level; see the discussion of ``candidate model''
and redundancy criterion in Section~\ref{sec:fdpll-me}.
On the other hand, two-level methods can take advantage of SAT-solving
technologies only by adapting them individually, both with respect to
theory and implementation.
A good example for this is \emph{lemma learning}, a key factor in
modern SAT-solving which helps to avoid repeating identical parts of a
refutation. It required some efforts to integrate lemma learning into Model
Evolution~\cite{Baumgartner:etal:ModelEvolutionLearning:LPAR:2006}
but, on the upside, led to a more powerful lemma learning mechanism.

Other differences between the IMs considered here would require a 
deeper technical treatment, something that is beyond the scope of this paper. 
See~\cite{Jacobs:Waldmann:ComparingIM:JAR:2007} for a comparison
of IMs with respect to simulations of refutations. 

In most applications, e.g., software verification, an efficient
treatment of equality by specialized inference rules is
mandatory. Fortunately, research on efficient equality reasoning in
IMs can capitalize on the results developed for the resolution
calculus over the last 20 years. Indeed, some promising approaches
along these lines have been developed for
Inst-Gen~\cite{Ganzinger:Korovin:InstEq:CSL:2004}, for Disconnection
tableaux~\cite{Letz:Stenz:EqualityDC:Tableaux:2002}, and for Model
Evolution~\cite{Baumgartner:Tinelli:ModelEvolutionCalculusEquality:CADE:2005}.
These approaches all employ ordering restrictions as pioneered for the
resolution calculus
(see~\cite{Nieuwenhuis:Rubio:ParamodulationTheoremProving:HandbookAR:2001}
for an overview).  How this is concretely realized and the discussion
of the differences is beyond the scope of the this paper.
A related but less developed topic is the integration of reasoning modulo
more general background theories, such as integer arithmetics or
theories of certain data structures (lists, arrays, sets, etc).
This is currently a hot topic, and only some initial results are 
available~\cite{Ganzinger:Korovin:ThInst:LPAR:2006,Baumgartner:Fuchs:Tinelli:MELIA:LPAR:2008}. 
One motivation for this stream of research to address a major weakness of
the prevailing approach to theory reasoning, \emph{Satisfiability
  Modulo Theories (SMT)}~\cite{Ranise:Tinelli:SMT:IEEE:2006}. 
To explain, current SMT systems are practically very successful for
quantifier-free (i.e.\ ground) input formulas. However, they do not natively
support quantifiers and resort to incomplete instantiation heuristics
for quantified formulas.
In contrast, IMs are devised as first-order logic calculi at the
outset and provide a systematic treatment of quantifiers. Equipping
IMs with theory reasoning could thus provide alternatives to
SMT. Under certain restrictions it is even possible to design
refutational complete calculi over, say, integer arithmetics.

A different line of research has only just begun, the combination of
instance based methods and resolution calculi. The motivation for that
is to combine their individual strengths in a single
framework. See~\cite{Baumgartner:Waldmann:MESUP:CADE:2009} for an
integration of Model Evolution and Superposition,
and~\cite{Lynch:McGregor:CombiningInstGenAndResolution:FroCoS:2009}
for an integration of Inst-Gen and Resolution.

Many of the calculi we discussed have been implemented, yielding insight into their practical applicability. 
For Hyper-Linking, there is a prover CLIN~\cite{clinPhd}, 
with improved versions CLIN-S (semantic restriction) \cite{clins} and CLIN-E (equality handling) \cite{cline}. 
For Disconnection Tableaux, there is a prover DCTP~\cite{dctp} 
featuring both equality handling and various refinements. 
The same is true for Inst-Gen, with the
iProver~\cite{Korovin:IPROVER:IJCAR:2008}, 
and the Model Evolution 
calculus, with prover Darwin~\cite{darwinArt}. DCTP, iProver and
Darwin regularly participate in the CASC  
competition.
 
Other basic research questions concern, for example, search space
improvements, implementation techniques, variants for deciding more
fragments of first-order logic than are currently known, better
understanding of theoretical properties, and clarifying the
relationships between IMs and other methods. 

\paragraph{Acknowledgements.} We thank the reviewers for their helpful
comments and suggestions.

\footnotesize 


\normalsize 


\begin{contact}  
Dr.\ habil.\ Peter Baumgartner\\
NICTA Canberra and Australian National University\\
7 London Circuit\\
Canberra ACT 2601\\
Australia \\
Email: Peter.Baumgartner@nicta.com.au
\end{contact}

\begin{vita}{PeterBaumgartner.jpg}
  \vitaauthor{Peter Baumgartner} is a Principal Researcher and
  Research Group Manager with NICTA, Australia's center of excellence
  in information and communication technology. (NICTA is funded by the
    Australian Government's \emph{Backing Australia's Ability} initiative.)
Before joining NICTA, he mainly worked at the University of Koblenz from 1990
until 2003, and at the Max-Planck-Institut for Computer
  Science in Saarbrücken from 2003 to 2005.
\end{vita}

\begin{vita}
\vitaauthor{Evgenij Thorstensen} is a graduate student at the
University of Oxford. He holds a master's   
degree in computer science from the University of Oslo.
\end{vita}

\end{document}